\documentclass[12pt]{iopart}
\pdfoutput=1 %for arXiv submission
\usepackage{iopams}
\usepackage{gensymb}
\usepackage{graphicx}
\usepackage{url}
\usepackage{comment}
\usepackage{tabularx}
\usepackage{array}
\usepackage[colorlinks]{hyperref}
\usepackage[dvipsnames]{xcolor}
\usepackage[symbol]{footmisc}
\usepackage[normalem]{ulem}
\usepackage{enumitem}

\renewcommand{\thefootnote}{\arabic{footnote}}

\def\lsim{\;\rlap{\lower 2.5pt
   \hbox{$\sim$}}\raise 1.5pt\hbox{$<$}\;}
\def\gsim{\;\rlap{\lower 2.5pt
 \hbox{$\sim$}}\raise 1.5pt\hbox{$>$}\;}

\hypersetup{
linkcolor=BrickRed
,citecolor=NavyBlue
,filecolor=Mulberry
,urlcolor=BrickRed
,menucolor=BrickRed
,runcolor=Green
}

\usepackage{aas_macros}
\usepackage{appendix}

\begin{document}

\title[Deep Learning Cosmic Ray Transport]{Deep Learning Cosmic Ray Transport from Density Maps of Simulated, Turbulent Gas}

\author[0000-0002-8366-2143]{Chad Bustard$^1$, John Wu$^{2,3}$}

\address{$^1$ Kavli Institute for Theoretical Physics, University of California - Santa Barbara, Kohn Hall, Santa Barbara, CA 93107, USA}
\address{$^2$ Space Telescope Science Institute, 3700 San Martin Dr, Baltimore, MD 21218}
\address{$^3$ Department of Physics \& Astronomy, Johns Hopkins University, 3400 N Charles St, Baltimore,
MD 21218}

\ead{bustard@ucsb.edu}

\begin{abstract}
The coarse-grained propagation of Galactic cosmic rays (CRs) is traditionally constrained by phenomenological models of Milky Way CR propagation fit to a variety of direct and indirect observables; however, constraining the fine-grained transport of CRs along individual magnetic field lines -- for instance, diffusive vs streaming transport models --  is an unsolved challenge. Leveraging a recent training set of magnetohydrodynamic turbulent box simulations, with CRs spanning a range of transport parameters, we use convolutional neural networks (CNNs) trained solely on gas density maps to classify CR transport regimes. We find that even relatively simple CNNs can quite effectively classify density slices to corresponding CR transport parameters, distinguishing between streaming and diffusive transport, as well as magnitude of diffusivity, with class accuracies between $92\%$ and $99\%$. As we show, the transport-dependent imprints that CRs leave on the gas are not all tied to the resulting density power spectra: classification accuracies are still high even when image spectra are flattened ($85\%$ to $98\%$ accuracy), highlighting CR transport-dependent changes to turbulent phase information. We interpret our results with saliency maps and image modifications, and we discuss physical insights and future applications.
\end{abstract}

\section{Introduction}
\label{sec:intro}
\subsection{Cosmic Ray Fundamentals}
Galaxies are complex, dynamic systems with collisional components such as gas reservoirs, and collisionless components that primarily interact through gravity (such as stars and dark matter). In a broad sense, the collisional composition of galaxies can be divided as follows: there is non-relativistic, typically ionized gas that we are most accustomed to thinking about, there are cosmic rays (CRs), which are high-energy, charged particles that travel through the Universe at close to the speed of light, and there are magnetic fields, which couple to both non-relativistic gas and relativistic CRs through electromagnetic forces. Each of these components shapes the gas flows that regulate star formation and the long-term evolution of galactic ecosystems, but there are significant unknowns with each component and their interplay.

In this paper, we concern ourselves with how CRs, on large scales\footnote{On ``large'' scales, typically greater than a parsec, CRs can be collectively described as a relativistic fluid instead of individual particles.}, transfer momentum and energy with the surrounding non-relativistic gas, which is very dependent on the highly uncertain and scale-dependent motion of the CR fluid relative to the background gas. On very large scales, we can average over many of the smaller, turbulent fluctuations in the Universe, and therefore average over the tangled magnetic field lines that guide and scatter CRs. This large-scale, \emph{coarse-grained} CR propagation is what is constrained by current state-of-the-art phenomenological models \cite{Hanasz2021}. These models make informed assumptions on the geometry of the Milky Way, inject CRs from their likely sources within the Milky Way disk, propagate CRs according to some plausible paradigms with tune-able parameters, and calculate a variety of direct and indirect CR indicators: for instance, gamma-ray emission from interactions between hadrons and CR protons, radio synchrotron emission from spiraling CR electrons, and secondary products of spallation, the direct collision of CRs with other gas particles in the Universe. The best configuration, which minimizes the differences between the model output and real observations, is one in which the coarse-grained transport of CRs is diffusive and energy-dependent \cite{Tjus2020, Hanasz2021}; however, these models  cannot tell us the zoomed-in, \emph{fine-grained} CR transport along individual field lines.

The fine-grained transport depends on the source and type of hydromagnetic waves that scatter and confine CRs (see e.g. \cite{ZweibelReview2017, CRReview2023} for recent reviews). If CRs scatter off compressible fast modes \cite{Yan2004} that are created by external turbulence and cascade down to the CR gyroscale ($\approx 0.1$ AU for a GeV CR in the Milky Way), then fine-grained CR transport is believed to be diffusive and predominantly parallel to the local magnetic field. On the other hand, if the scattering waves are created by the CRs themselves through the so-called ``streaming instability" \cite{Wentzel1968, Kulsrud1969}, then CR transport is referred to as ``streaming", which can be a mixture of field-aligned diffusion and additional field-aligned advection at the Alfv\'{e}n speed $v_{A} = B/\sqrt{4\pi\rho}$, where $B$ is the magnetic field strength and $\rho$ is the gas density.    

For a multitude of reasons, identifying the true, fine-grained CR transport mode is crucial. Despite representing only a billionth of all particles in the Milky Way, CRs on a whole have as much energy as normal, non-relativistic gas \cite{Boulares1990}, and it is clear from a veritable explosion of work in the last decade (e.g. \cite{Simpson2016, Wiener2017, Pfrommer2017, Ruszkowski2017, Buck2020, Hopkins2020CRs, JiCRHalos2020, Bustard2021,  Huang2022, CGMReview2023, CRReview2023}) that the content of CRs in various astrophysical environments and the dynamical and thermodynamical influence of CRs on the surrounding gas sensitively depend on this transport. For example, in simulations of the Large Magellanic Cloud (LMC), a neighboring satellite galaxy of the Milky Way, if one allows CRs to stream, the galaxy remains largely intact over long periods of time as CRs easily lose pressure and escape the galaxy; however, if one replaces streaming with a small diffusivity instead, CRs build up a large pressure gradient in the galaxy and expel gas in a large-scale ``galactic wind" \cite{Bustard2020}. To reconcile observed galaxy gas contents with simulations and to predict the future gas content of galaxies, including those like the LMC that will eventually collide with the Milky Way, we need to know the true mixture of streaming vs diffusive CR transport.

So how can we determine whether CR transport is streaming or diffusive? One intriguing approach, motivated by very recent results, is to use the distinct, transport-dependent imprints that CRs leave on their surroundings. The basis for this idea is that diffusing and streaming CRs interact with gas fluctuations in fundamentally different ways. Denoting the CR pressure as $P_{\rm CR}$, diffusing CRs have flux $F_{\rm CR} \propto \nabla P_{\rm CR}$, which introduces a CR perturbed force that is proportional to velocity and creates a $\pi/2$ phase shift between CR pressure and gas density perturbations. Akin to a damped harmonic oscillator, this damps the waves \cite{ptuskin81}, leading to CR acceleration. Streaming CRs, with flux $F_{\rm CR} \propto P_{\rm CR}$, do not induce such a phase shift and have decreased acceleration rates \cite{BustardOh2022_reacceleration}, but they transfer energy to the gas and can drive unique instabilities, for instance of acoustic waves in highly magnetized plasmas \cite{Begelman1994} as seen recently in idealized 1D simulations \cite{Tsung2021_staircase, Quataert2022StreamingCRs}.

Transport-dependent impacts on gas are also readily apparent in fully 3D simulations. Bustard and Oh 2023 \cite{BustardOh2023_arxiv} simulated CR-gas interactions in subsonic, compressive turbulence and found that turbulent energy spectra change dramatically depending on CR transport mode, with all other variables (CR pressure, stirring rate, etc.) held fixed. Namely, when CR diffusion dominates, CRs take energy from the gas and gain energy themselves\footnote{The subsequent CR energy gain is known as turbulent reacceleration \cite{Ptuskin1988, Brunetti2011}.}, introducing cut-offs and new slopes to kinetic energy spectra compared to a no-CR case. CR streaming alters this damping \cite{BustardOh2022_reacceleration}, affecting turbulent spectra, gas thermodynamics, and density structures in a distinctly different manner \cite{BustardOh2023_arxiv}. 

\subsection{Goals of this Work}

Overall, the Bustard and Oh 2023 simulation suite provides terabytes of unstructured gas density images stemming from otherwise identical simulations but with different CR transport assumptions. The primary goal of this paper is to explore whether deep convolutional neural networks (CNNs) can learn to accurately predict the CR transport encoded in each image, and more importantly, whether subsequent network interpretation using image manipulation and saliency maps can help illuminate the most salient, distinguishing features of gas density maps. To that end, our aim is to train a CNN to high enough accuracy to enable useful interpretation and reveal new insights into how CRs affect their surroundings. In the following exploratory analysis, we use density slices from the Bustard and Oh 2023 simulation suite as our training and validation data, and we train and fine-tune CNNs using PyTorch \cite{NEURIPS2019_9015}, a popular and open-source Python-based deep learning framework, to classify density images into one of five sets of simulations, varying only in the CR transport model assumed.

A second question, which we largely defer to future work, is whether these neural networks trained on simulations can be used accurately in production with real observations as inputs. This possibility is quite interesting: instead of requiring the full, and expensive to acquire, multi-wavelength observations input into phenomenological models \cite{StrongReview2007, Hanasz2021}, all one theoretically needs are images of HI (neutral hydrogen) density obtained from a high-resolution survey. Given the highly idealized nature of even these state-of-the-art turbulence simulations and the significant uncertainty as to whether idealized simulations capture the density map differences of real observations\footnote{For example, simulations may not be sufficiently converged with resolution, or other potentially dominant complications due to dust or multiphase gas can exist.}, it is premature to conduct a full study of this possibility. Instead, we discuss this domain adaptation in Section \ref{sec:conclusions} and briefly show that, in light of our results in Section \ref{sec:classify}, a universal challenge complicating all astronomical analyses is especially relevant here: the depth of 3D structures that forms a 2D image is highly uncertain, and varying this depth significantly changes the accuracy of our network.

The outline of this paper is as follows. In Section~\ref{sec:setup}, we recap the simulations of Bustard and Oh 2023, describe the data preprocessing steps we take, and outline the basic components of our CNN architecture. In Section~\ref{sec:classify}, we present our classification results, first on the entire fiducial dataset spanning all five classes. We also present our interpretation of these results using saliency maps, and we probe the limitations of CNNs further by flattening the power spectra of our images and Gaussian filtering our input images. In \ref{sec:appendix}, we explore how well a network trained on single-cell-thick slices of a density cube can classify more realistic projections over multiple cells, highlighting the need for additional training on a larger, realistic, and more diverse image set. We conclude in Section~\ref{sec:conclusions}.  

\emph{Code availability: } All Python scripts used in this work are hosted at \url{https://github.com/bustardchad/ML_Turb}, including descriptive Jupyter notebooks.

\emph{Data availability: } A subset of data is hosted through the Harvard Dataverse at \url{https://doi.org/10.7910/DVN/WBY5CX}

\section{Training Data and CNN Architecture}
\label{sec:setup}
\subsection{Simulation Sets and Labels}

The data for this project comes from the Bustard and Oh 2023 turbulent box simulation suite, and we encourage readers to see Section 2 of \cite{BustardOh2023_arxiv} for further details. As a brief recap, these simulations are all run using the Athena++ magnetohydrodynamics (MHD) code \cite{AthenaRef} with an additional module that includes CRs as a relativistic fluid with adiabatic index $\gamma_{c} = 4/3$ and with energy and flux coupled to the normal MHD equations (see \cite{JiangCRModule} for more details), including terms for field-aligned CR streaming and diffusion. The non-relativistic gas is treated as an isothermal (constant temperature) fluid with adiabatic index $\gamma_{g} = 1$. Purely compressive turbulence is stirred according to an Ornstein-Uhlenbeck random process \cite{Uhlenbeck1930} centered on scale $L$ in a cubic box of size $(2L)^{3}$, leading to an effective turbulent outer scale of $L_{0} \approx 2L/3$. In each simulation, the turbulent stirring rate, correlation time, etc. are all kept fixed, and in this study, we focus on the simulations that, in absence of CRs, produce a sonic Mach number $M_{s} \sim 0.15$, defined as the ratio of the turbulent velocity $v$ to the gas sound speed $c_{s}$. The initial composite mixture of gas, magnetic fields, and CRs is constant for all simulations, and has initial gas-to-magnetic pressure ratio $\beta = P_{g}/P_{B} \sim 10$ and CR-to-gas pressure ratio $\eta = P_{CR}/P_{g} \sim 1$, except the MHD-only simulation, for which there are no CRs. The initial magnetic field configuration is in the $\hat{x}$-direction. The stirring generates sub- to trans-Alfv\'{e}nic ($\mathcal{M}_{A} = v/v_{A} < 1$) turbulence, and with purely compressive forcing (rather than solenoidal forcing), there is no appreciable amplification of the magnetic field.

\begin{table}
  \centering
  \begin{tabular}{l|r}
    Class Name & CR Transport Parameters \\
    \hline 
    \texttt{MHD} & No CRs \\
    \texttt{CR{\_}Advect} & $\kappa \sim 0$  \\
    \texttt{CR{\_}Diff{\_}Fiducial} & $\kappa = \kappa_{f} \sim 0.1 L_{0}c_{s}$  \\
    \texttt{CR{\_}Diff100} & $\kappa = 100\kappa_{f} \sim 10 L_{0}c_{s}$  \\
    \texttt{CR{\_}withStreaming} & $\kappa = \kappa_{f}$ + streaming \\
    
  \end{tabular}
    \caption{Class labels and image set names. For each simulation, the following are constant: $\beta = P_{g}/P_{B} \sim 10$, $\eta = P_{CR}/P_{g} \sim 1$ (except $\eta = 0$ for the \texttt{MHD} class), the box size $(2L)^{3}$, the outer eddy size $L_{0} \sim 2L/3$, and all stirring parameters (see \cite{BustardOh2023_arxiv}).}
  
\label{table1}
\end{table}

Importantly, compressive motions transfer kinetic energy to the CRs. Prior analyses showed that the rate of this CR energization depends on CR transport mode and the gas-to-magnetic pressure ratio $\beta$ \cite{BustardOh2022_reacceleration}, while Bustard and Oh 2023 showed that this transport-dependent energy transfer affects the turbulent kinetic energy cascade and the spectra of gas density structures. Namely, there is a ``sweet-spot" CR diffusion coefficient $\kappa \sim 0.1 L_{0} c_{s}$ where CRs most severely damp turbulent fluctuations, leading to a steeper spectral slope and a lack of small-scale power in the cascade compared to simulations with non-optimal CR diffusivity. Our simulation classes are summarized in Table \ref{table1}, with example images shown in Figure \ref{fig:example_images} and with expanded descriptions of each class given below:

\begin{figure*}
\centering
\includegraphics[width=0.99\textwidth]{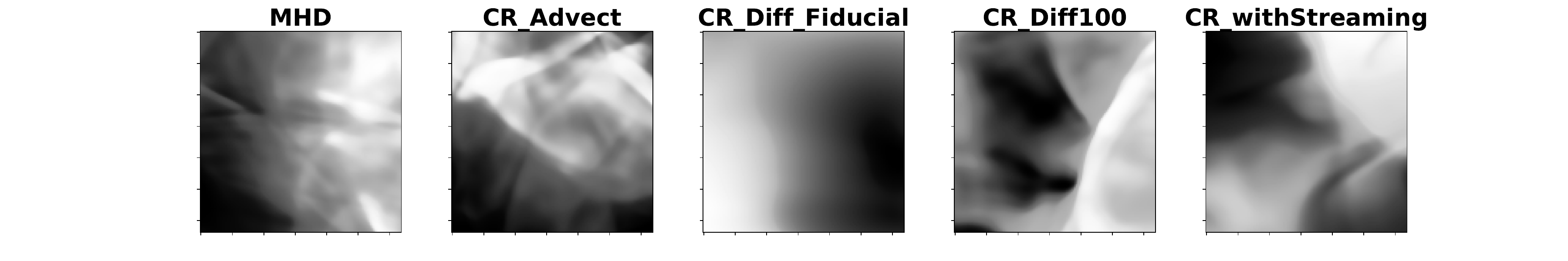}
\includegraphics[width=0.99\textwidth]{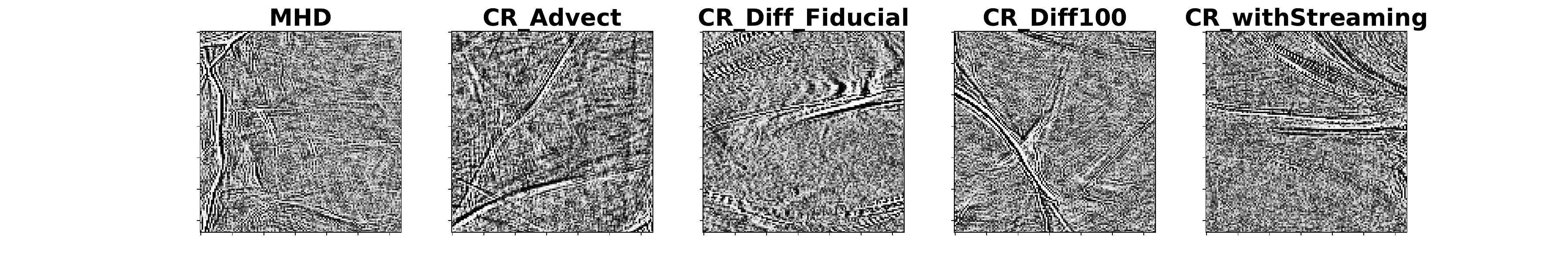}
\caption{Example gas density images for each class. The top row shows unaltered images from the \texttt{Full Power} set, while the bottom row shows another, distinct set of images after their power spectra have been flattened (from the \texttt{Flattened Power} set). Note in the top row some of the major differences between \texttt{CR{\_}Diff{\_}Fiducial}, which shows very smooth transitions between over- and under-dense gas, and e.g. \texttt{MHD}, which has sharp transitions and more small-scale structure.} 
\label{fig:example_images}
\end{figure*}

\begin{figure}
\centering
\includegraphics[width=0.6\textwidth]{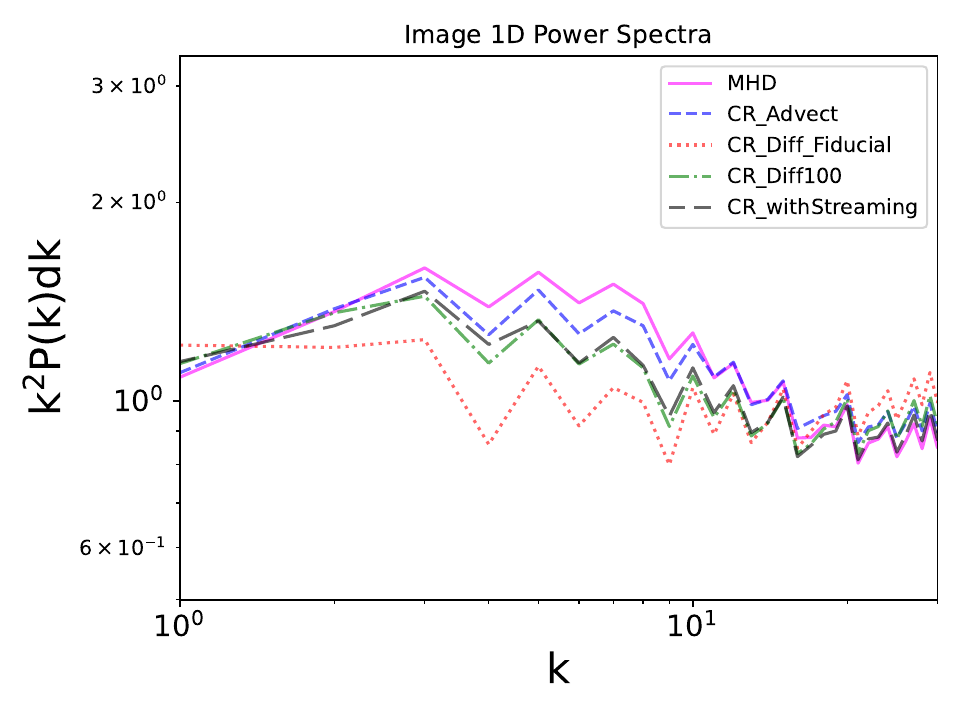}
\caption{Average 1D gas density power spectra (further multiplied by $k^{2}$) over a batch of 512 images from the test set, showing noticeable spectral differences between \texttt{CR{\_}Diff{\_}Fiducial} and other sets but relatively small differences between the other CR sets. } 
\label{fig:original_spectra}
\end{figure}

\begin{itemize}
\item \texttt{MHD}, a no-CR ($\eta = P_{CR}/P_{g} = 0$) control case. Turbulence in this case is entirely formed by MHD effects. On well-resolved scales (wave numbers k $< 10-20$), there is significant power, as CRs are not present to play a damping role. 
\item \texttt{CR{\_}Advect}, with roughly equal pressure contributions from CRs and gas ($\eta = P_{CR}/P_{g} \sim 1$), but with no CR diffusivity ($\kappa$ = 0). We refer to this case as \texttt{CR{\_}Advect} because CRs only \emph{advect} with the gas. The major physical difference, then, is that the composite CR and gas mixture has an effective equation of state somewhere between that of the isothermal, non-relativistic gas, where pressure and density are related by $P \sim \rho$, and a relativistic gas, where $P \sim \rho^{4/3}$. This slightly affects the compressibility of the gas and the resulting density images, but as we see from Figure \ref{fig:original_spectra}, the 1D density power spectrum is very similar to the \texttt{MHD} case, suggesting these classes will be hard to disentangle. 
\item \texttt{CR{\_}Diff{\_}Fiducial}, where CRs are present with a fiducial diffusivity $\kappa_{f} \sim 0.1 L_{0} c_{s}$. This diffusivity optimizes the energy transfer between gas fluctuations and CRs, leading to the most significant damping, especially of intermediate wavenumber fluctuations (see Figure \ref{fig:original_spectra}).
\item \texttt{CR{\_}Diff100}, where $\kappa \rightarrow 100 \kappa_{f}$. With such a high diffusivity, CRs flow over gas fluctuations so quickly that they don't damp them as effectively. This leads to gas density power spectra intermediate between the \texttt{MHD} and \texttt{CR{\_}Diff{\_}Fiducial} cases.
\item \texttt{CR{\_}withStreaming}, which includes both fiducial diffusion and CR streaming. This case is of particular interest because of the unique changes that streaming imparts on the turbulence. Instead of CRs taking energy from gas motions and keeping it, the amount of turbulent damping is lower, and much of the energy that CRs receive is deposited back into the gas as heat at scales far larger than the typical dissipation scale. The resulting turbulent energy spectrum does not display such an obvious cut-off or change in spectral slope, but is instead suppressed almost uniformly across all scales (see Figure 8 in \cite{BustardOh2023_arxiv}). If one scales and normalizes the resulting density images, as we do in this work, the streaming spectrum is almost exactly the same as the \texttt{MHD}, \texttt{CR{\_}Advect}, and \texttt{CR{\_}Diff100} spectra, as we see in Figure \ref{fig:original_spectra}.
\end{itemize}

\subsection{Training Data and Preprocessing}

\begin{figure}
\centering
\includegraphics[width=0.6\textwidth]{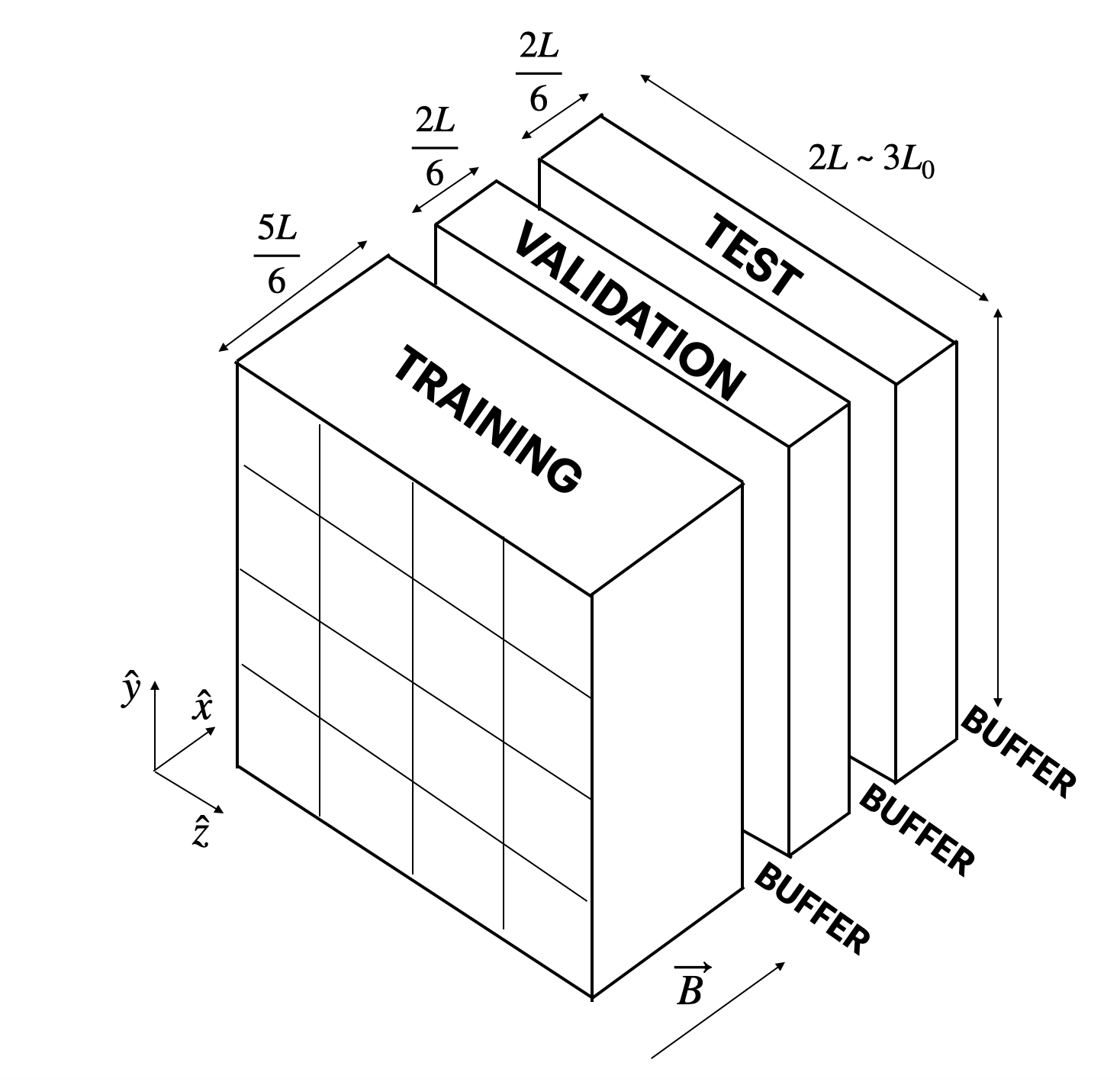}
\caption{Cartoon showing how each data snapshot with total volume of $(2L)^{3}$ is split into training, validation, and test sets. Our training set occupies a volume of ($2L \times 2L \times 5L/6$), while our validation and test sets are 2.5 times smaller with volumes of ($2L \times 2L \times 2L/6$). Each set is spatially separated from other sets by a buffer of width $L/6$ in the $\hat{x}$ direction to decrease correlations. Within each set, images of size $128 \times 128$ cells are created by slicing in the y-z plane, perpendicular to the initial magnetic field $\vec{B}$}. 
\label{fig:box}
\end{figure}

Each simulation snapshot contains $512^{3}$ cells, and for each simulation class, we utilize 6 time snapshots temporally separated by at least an (outer-scale) eddy turnover time to ensure snapshots are sufficiently uncorrelated. Our process for splitting these data cubes into images largely follows that of \cite{Peek2019}, who similarly analyzed MHD turbulence simulations to classify sub-Alfv\'{e}nic vs super-Alfv\'{e}nic regimes. To create our image sets, we cut the box into 512 one-cell-thick slices of dimension $512 \times 512$ in the y-z plane (transverse to the initial guide magnetic field), and from each of those slices, we create 16 images of size $128 \times 128$. 

Since eddies in magnetized turbulence are elongated along the background magnetic field direction $\hat{x}$ \cite{GoldreichSridhar1995}, slices in the x-z or x-y planes will contain imprints of the original magnetic field. The extent of this eddy anisotropy, which imprints on gas density images, would not bias our analysis because it is indeed a physical outcome of our simulations that start from exactly the same initial conditions. Nevertheless, to ensure that this anisotropy, which is dominantly caused by magnetic field effects rather than CR effects, is not a feature that our network can use to distinguish different CR transport modes, we follow \cite{Peek2019} and slice across the magnetic field axis, forcing the network to find other distinguishing image characteristics more likely to be caused by CRs.

We also note here that, by taking a one cell thick slice, we are effectively integrating over structures on the scale of a simulation cell width. How well this reflects the projection depth of a real astronomical image is complicated and very dependent on the astrophysical environment, as we discuss in Section \ref{sec:appendix}. In this paper, however, we focus on how neural network interpretation can help us derive new insights from simulations, and our slicing choice is sufficient.

From here, we must be careful to spatially separate the training, validation, and test slices so that structures correlated in the $\hat{x}$-direction do not bleed between sets and introduce correlations. To decrease the chances of this, we put spatial buffers between the training, validation, and test sets. The training set occupies a width of $5L/6$, followed by a buffer of width $L/6$, then a validation set of width $2L/6$, then a buffer, then a test set of width $2L/6$, then a final buffer (Figure \ref{fig:box}). Our results do not seem particularly sensitive to buffer size, except in the case with \emph{no} buffer where our saliency maps (Section \ref{sec:interp}) were dominated by pixel-scale regions, indicative of the CNN ``memorizing" regions of the training set that were correlated with the validation set. This problem was particularly evident when we created training and validation sets by randomly choosing slices from the 3D data volumes; in this case, structures very much span across images from both sets, leading to a network with no ability to generalize to unseen data. An alternative way to split training, validation, and test sets could be to separate them temporally. For instance, training data could comprise snapshots 1-4, validation snapshot 5, and test snapshot 6; however, this means images within each set are not well-separated in time. By instead creating sets that span across all times available, our network is trained, validated, and tested on more diverse manifestations of the turbulent gas-CR interactions.

Within each set, we fiducially keep half of the images; we primarily do this to keep our dataset sizes small enough to be loaded into RAM, but this can also help decrease spatial correlations within each set. To further decrease correlations, we randomly flip the images both horizontally and vertically, each with a $50\%$ probability. In all, our fiducial training, validation, and test sets contain $\sim$ 10,000, 4,000, and 4,000 images per class (50,000, 20,000, and 20,000 total). Each image is then preprocessed as follows:

\begin{enumerate}
    \item The density is logarithmically scaled. Because turbulent density probability distribution functions (PDFs) are roughly lognormal, this scaling brings out more features that would otherwise be sub-dominant compared to the most dense regions.

    \item The images are histogram equalized using the exposure method from scikit-image \cite{scikit-image} such that image pixels have a roughly equal distribution of values from 0 to 1. As noted in \cite{Peek2019}, this step is a common preprocessing step used to optimize the CNN, but it eliminates density PDF information from our images. With one of our goals to see if neural networks can find information \emph{beyond} PDF and spectral information, this is perfectly acceptable.
    
\end{enumerate}

Images processed in this way encompass our \texttt{Full Power} image set in that they retain spectral information. As in \cite{Peek2019}, we also create a \texttt{Flattened Power} image set with no spectral information by applying a fast Fourier transform to each image and setting the Fourier power to unity; this happens in between steps (i) and (ii) above. In this case, the neural network is left to only distinguish image classes based on image \emph{phase} information. 

\subsection{Neural Network Details}

Convolutional neural networks (CNNs) are a powerful deep learning architecture for computer vision, and as they have now been employed for various tasks in astrophysics, we do not give a long introduction to them here. Instead, we describe the key components and our choices for number of layers, number of trainable parameters, etc., and refer the audience to a recent review of deep learning in astrophysics (i.e. \cite{2023PASA...40....1H}).

The building blocks of CNNs are convolution layers, pooling layers, and fully connected layers, followed in this classification application by a softmax output layer that generates a probability of the input image belonging to each class. For the results presented here, we use 4 convolutional layers, each followed by batch normalization and SiLU activations. These layers take an input array (in the first layer, this input is the $128 \times 128$ image), and apply filters to sub-patches of the input, thereby generating many convolutions of the input. Batch normalization then normalizes the layers' inputs by re-centering and re-scaling them, making training faster and more stable. 

After these 4 convolution layers, we apply a pooling layer and then apply dropout with 25\% probability. We then flatten the output before sending it to a fully connected layer, where all neurons from the previous layer are connected to all neurons of the next layer. The output of this final layer, after going through a softmax activation, is a vector of probabilities that the input image corresponds to each class. For our full model with 5 classes, this vector has a length of 5, and when we make our final prediction of which class the image came from, we choose the class with highest probability.

All-in-all, this fiducial network, containing 29,749 trainable parameters, is appropriately sized for our dataset of $\sim 50,000$ training images; adding more layers leads to overfitting (high accuracy on training data but poor generalization to unseen data), while decreasing the number of layers leads to underfitting (poorer accuracy on training data). Networks for the \texttt{Full Power} and \texttt{Flattened Power} datasets are trained for 40 and 25 epochs, respectively, beyond which the models begin to overfit.

To speed up training, we employ mini-batch gradient descent with 64 images per batch. Weights and biases are updated during training using the AdamW optimizer \cite{AdamWOptim} with weight decay of $10^{-4}$ and a learning rate of $10^{-3}$ in the \texttt{Full Power} case and $5 \times 10^{-4}$ in the \texttt{Flattened Power} case. This gradient descent method, which is a modification to the popular Adam method \cite{AdamOptim}, decouples weight decay from the gradient update steps and improves generalization performance. The loss function that AdamW seeks to minimize is the cross-entropy loss between the predicted distribution and the true class distribution. 

All of the above choices were motivated by a limited, manual hyperparameter study, where we varied the learning rate between $10^{-3}$ and $10^{-4}$, the batch size between 8 and 256, and the dropout fraction between 0 and 0.5. We also tested the ReLU activation function instead of the SiLU activation function in hidden layers of our network, ultimately finding insignificant differences in training time and accuracy. The continuously differentiable SiLU behaves similarly to other activation functions (e.g., \cite{SiLuPaper,swish,Mish}), which are robust against the ``dying neuron'' problem with ReLU, and have been shown to improve performance in astronomical tasks (e.g., \cite{Guo+22,Wu+2022}).

\section{Results}
\label{sec:classify}

\subsection{Full Spectra Results}

\begin{table}
\begin{center}
\begin{tabular}{lrrrr}
\hline
Data set &  Accuracy &  Precision &    Recall &  F1-Score \\
\hline\hline
\texttt{Full Power}\\
\hspace{1em}\texttt{MHD}                &      95.5 &       99.5 &    77.9 &      87.4 \\
\hspace{1em}\texttt{CR\_Advect}         &     95.6 &       82.8 &    98.4 &      89.9 \\
\hspace{1em}\texttt{CR\_Diff\_Fiducial} &     99.2 &       96.0 &   100.0 &      98.0 \\
\hspace{1em}\texttt{CR\_Diff100}        &     92.0 &       77.7 &    84.2 &      80.8 \\
\hspace{1em}\texttt{CR\_withStreaming}  &     94.2 &       89.4 &    80.8 &      84.9 \\
\hline
\texttt{Flattened Power}\\
\hspace{1em}\texttt{MHD}                &     88.5 &       99.6 &    42.8 &      59.9 \\
\hspace{1em}\texttt{CR\_Advect}         &     96.7 &       96.7 &    86.4 &      91.2 \\
\hspace{1em}\texttt{CR\_Diff\_Fiducial} &     97.9 &       99.7 &    89.8 &      94.5 \\
\hspace{1em}\texttt{CR\_Diff100}        &     90.2 &       72.5 &    82.4 &      77.1 \\
\hspace{1em}\texttt{CR\_withStreaming} &     85.7 &       58.7 &    96.2 &      72.9 \\
\hline
\end{tabular}
\caption{A table of metrics comparing classification results on \texttt{Full Power} simulations and on \texttt{Flattened Power} simulations. The accuracy, precision, recall, and F1 scores are shown as percentages. \label{tab:metrics}}
\end{center}
\end{table}

\begin{figure*}
\centering
\includegraphics[width=0.90\textwidth]{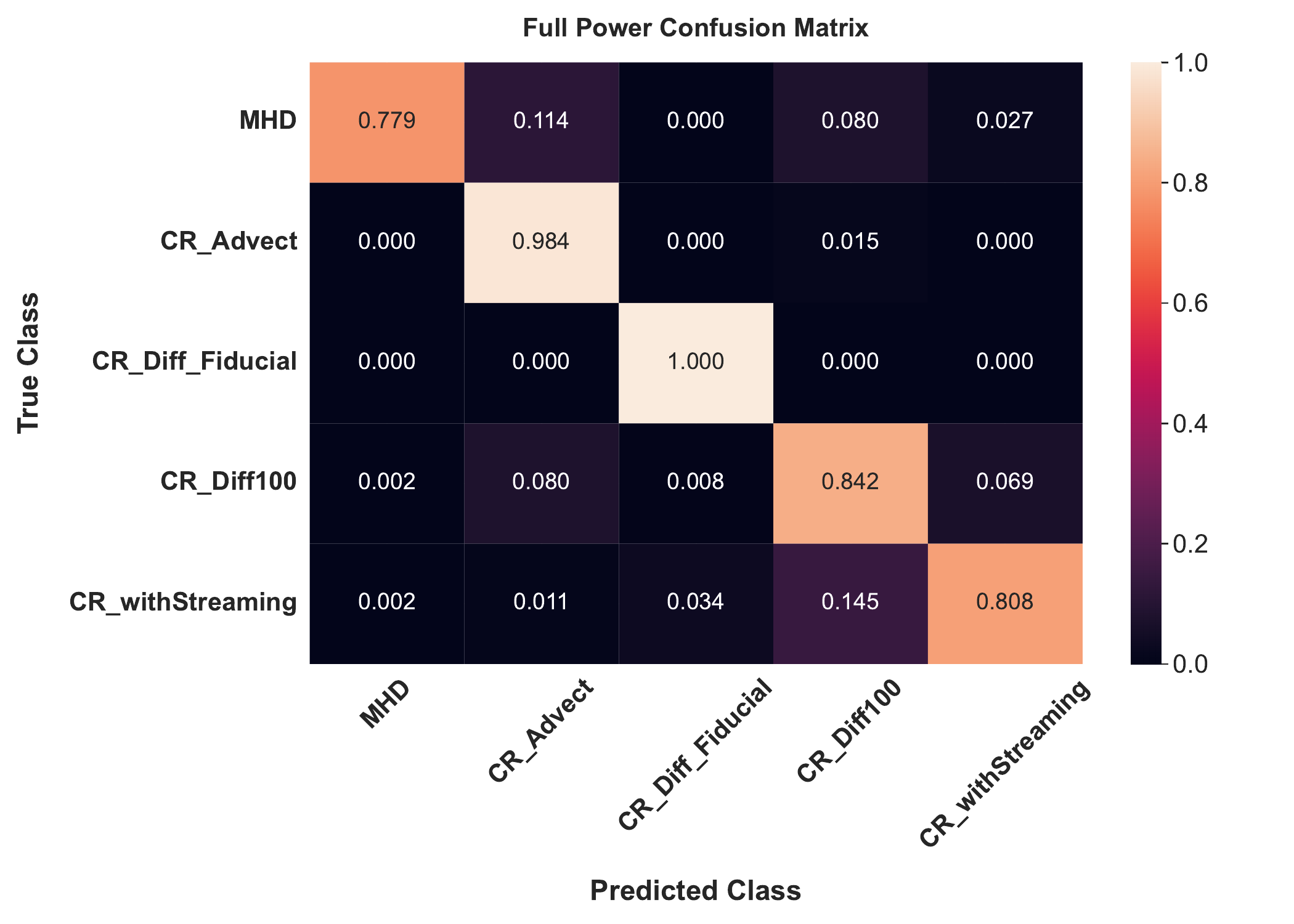}
\caption{Confusion matrix for the \texttt{Full Power} set of test images, showing the fraction of images in each predicted class vs their true class. Fractions are obtained by taking the raw number of images in each category and dividing by 3981, the number of images per class. Class accuracies range from $92.0\%$ for the \texttt{CR{\_}Diff100} class to $99.2\%$ for the \texttt{CR{\_}Diff{\_}Fiducial} class, which also achieves a perfect recall (see Table \ref{tab:metrics}). The average recall is $\approx 88.3 \%$, brought down significantly by the \texttt{MHD} images.} 
\label{fig:confusion_full_power}
\end{figure*}

In Table~\ref{tab:metrics}, we display several commonly used metrics for multi-class machine learning problems, which depend on the number of true positives (TP), true negatives (TN), false positives (FP), and false negatives (FN).
The accuracy is defined as (TP + TN) / (TP + TN + FP + FN), the precision is defined as the TP / (TP + FP), the recall is defined as the TP / (TP + FN), and the F1 score is the harmonic mean between the precision and recall. The precision can be considered a measure of ``purity'' while the recall can be thought of as ``completeness'' for CNN predictions. Figure \ref{fig:confusion_full_power} shows the confusion matrix for the \texttt{Full Power} test set, created with scikit-learn \cite{scikit-learn} and showing raw counts of images with predicted vs true labels. 

The accuracy for each class is quite high, ranging from $92.0\%$ for \texttt{CR{\_}Diff100} to $99.2\%$ for \texttt{CR{\_}Diff{\_}Fiducial}. Precision and recall vary more significantly, leading to generally lower F1 scores. For instance, recall ranges from $\sim 78 \%$ for the \texttt{MHD} class to $100 \%$ for the \texttt{CR{\_}Diff{\_}Fiducial} class, but especially for the \texttt{MHD} class, low recall is compensated by high precision ($99.5\%$ for \texttt{MHD}). This tendency for a CNN to trade recall for precision or vice versa is a common and nonlinear behavior; therefore, it is critical to report multiple summary statistics and combined metrics like the F1 score\footnote{In practice, one could change the softmax function in our network to include a tunable parameter $\alpha$, i.e. softmax($\alpha$,z) = exp(-$\alpha$z)/$\sum_{z}$exp(-$\alpha$z), and evaluate the F1 score on the validation set over a grid of $\alpha$ values to find the optimal trade-off between precision and recall averaged over all classes. However, this does not guarantee that one will find an $\alpha$ that simultaneously maximizes the F1 score for each individual class, and we proceed with a default value of $\alpha=1$.}.

The average recall is $88.3 \%$, brought down most significantly by the \texttt{MHD} case, which is confused for other classes $22 \%$ of the time. In particular, \texttt{MHD} is incorrectly labeled as \texttt{CR{\_}Advect} $11.4 \%$ of the time and as \texttt{CR{\_}Diff100} $8.9 \%$ of the time. These confusions make physical sense: advecting CRs do not sap any energy from turbulent fluctuations. Instead, the inclusion of CRs only changes the composite gas+CR equation of state because relativistic CRs have a $\gamma = 4/3$ adiabatic index instead of a $\gamma = 5/3$ index for a non-relativistic gas. The resulting images are quite comparable. CRs with large diffusivity (\texttt{CR{\_}Diff100}) do not appreciably interact with fluctuations either; their fast transport (on a short diffusive timescale $\tau_{\rm diff} \sim L^{2}/\kappa$) means they pass over the flow too quickly for eddies to interact with the CRs during an eddy turnover time $\tau_{\rm eddy} \sim L/v$, essentially leaving turbulence and the resulting density image untouched. 

One might wonder whether the MHD simulations are necessary when, for instance, we know quite certainly that MHD-only is a poor approximation in the Milky Way interstellar medium where $P_{\rm CR} \sim P_{g}$ \cite{Boulares1990}. To test this scenario, we also trained and fine-tuned networks on only the 4 CR classes. For brevity, we do not show the resulting confusion matrix, but we note that we obtained very similar statistics with only slightly boosted F1 scores. This implies that the full, 5-class network is capable of discriminating between CR classes, despite being presented with confusing MHD images. 

Most impressively, the \texttt{CR{\_}Diff{\_}Fiducial} and \texttt{CR{\_}withStreaming} classes, which differ only in that CR streaming is included in addition to fiducial diffusivity, are well-distinguished, with the network achieving $94.2\%$ accuracy on \texttt{CR{\_}withStreaming} and only rarely ($3.4 \%$ of the time) confusing \texttt{CR{\_}withStreaming} for the \texttt{CR{\_}Diff{\_}Fiducial} class. Instead, \texttt{CR{\_}withStreaming} is confused for \texttt{CR{\_}Diff100} $\approx 14.5 \%$ of the time, likely because additional CR streaming means CRs are propagating faster along field lines, somewhat akin to faster diffusion. How fast is streaming transport? In these simulations, turbulence is sub-Alfv\'{e}nic, meaning the characteristic CR transport speed (the Alfv\'{e}n speed $v_{A}$) is faster than the turbulent velocity\footnote{In fact, this characteristic transport speed is a lower limit only realized when CRs are well-coupled to Alfv\'{e}n waves, which only occurs when CR pressure gradients are aligned with the magnetic field. In turbulence, these vectors are frequently misaligned \cite{BustardOh2022_reacceleration}, leading to macroscopic CR decoupling from waves and the so-called ``bottleneck effect", where CRs free-flow at relativistic speeds and develop a flat pressure gradient unable to transfer momentum and energy to the gas \cite{Skilling1971, Wiener2017}.}, meaning the CR transport time $\tau_{\rm stream} \sim L/v_{A} < L/v \sim \tau_{\rm eddy}$, similar to the fast diffusion case where $\tau_{\rm diff} \sim L^{2}/\kappa < L/v \sim \tau_{\rm eddy}$. Previous work \cite{BustardOh2023_arxiv}, however, hints that additional discriminating information will be present, such as transport-dependent ratios of compressive vs solenoidal motions in the gas, which likely accounts for some of the accurate differentiation between \texttt{CR{\_}withStreaming} and other CR classes. In all, the accuracies we obtain with our relatively simple network are high enough to continue with network interpretation, and misclassification trends shown by our confusion matrices already reveal distinct differences between some classes and interesting morphological overlaps between others.

\subsection{Network Interpretation}
\label{sec:interp}

\setkeys{Gin}{draft=false}
\begin{figure*}
\centering
\includegraphics[width=0.85\textwidth]{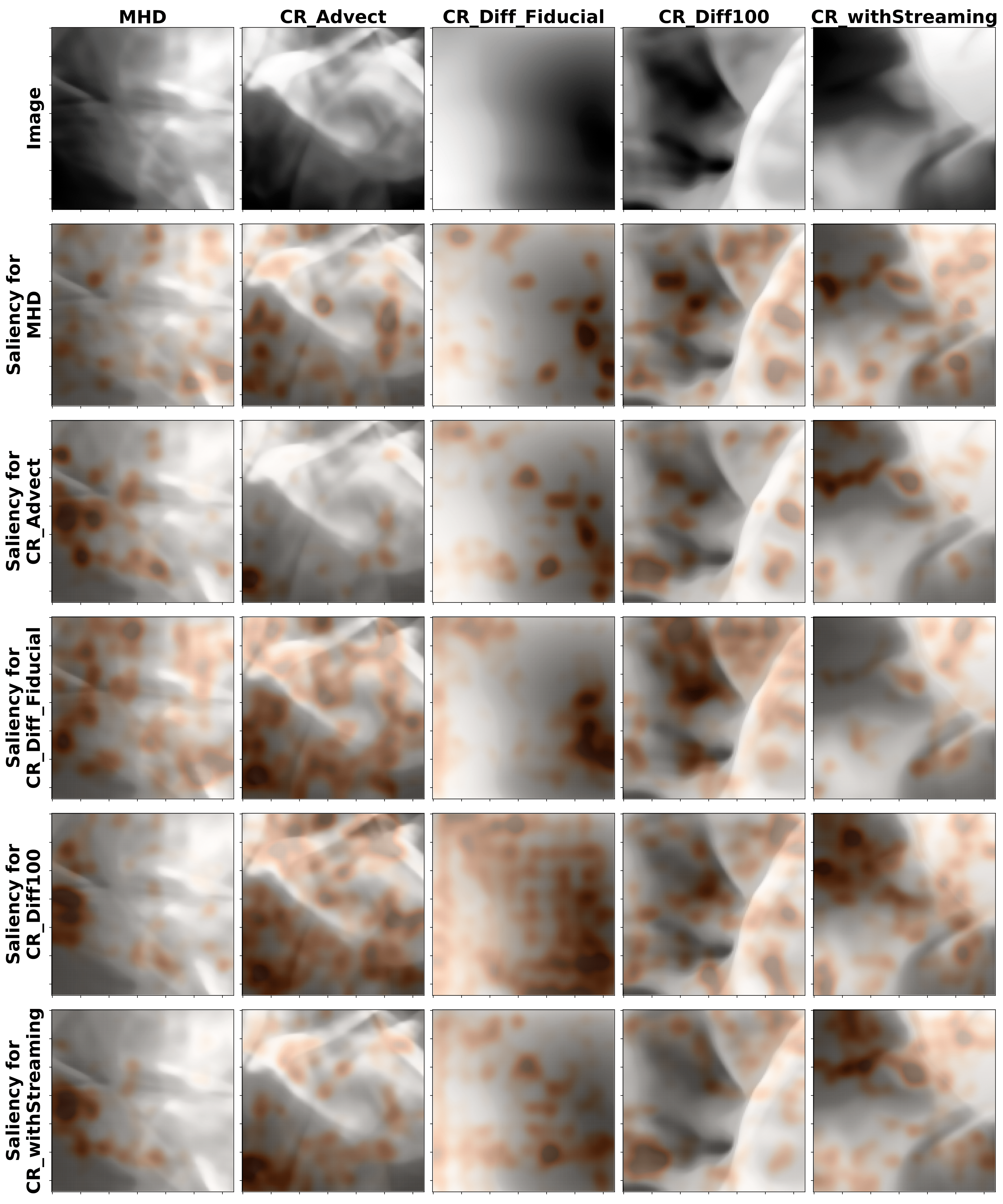}
\caption{Saliency maps for the example \texttt{Full Power} images of Figure \ref{fig:example_images}. Each row shows the same image from a given class, while each column shows activations for each of the 5 classes when the trained network is presented with that image. Activations are Gaussian smoothed instead of shown pixel-by-pixel, and they are normalized to the range [0,1], which amplifies otherwise small activations for some classes.}
\label{fig:saliency_colormesh}
\end{figure*}

\setkeys{Gin}{draft=false}
\begin{figure*}
\centering
\includegraphics[width=0.85\textwidth]{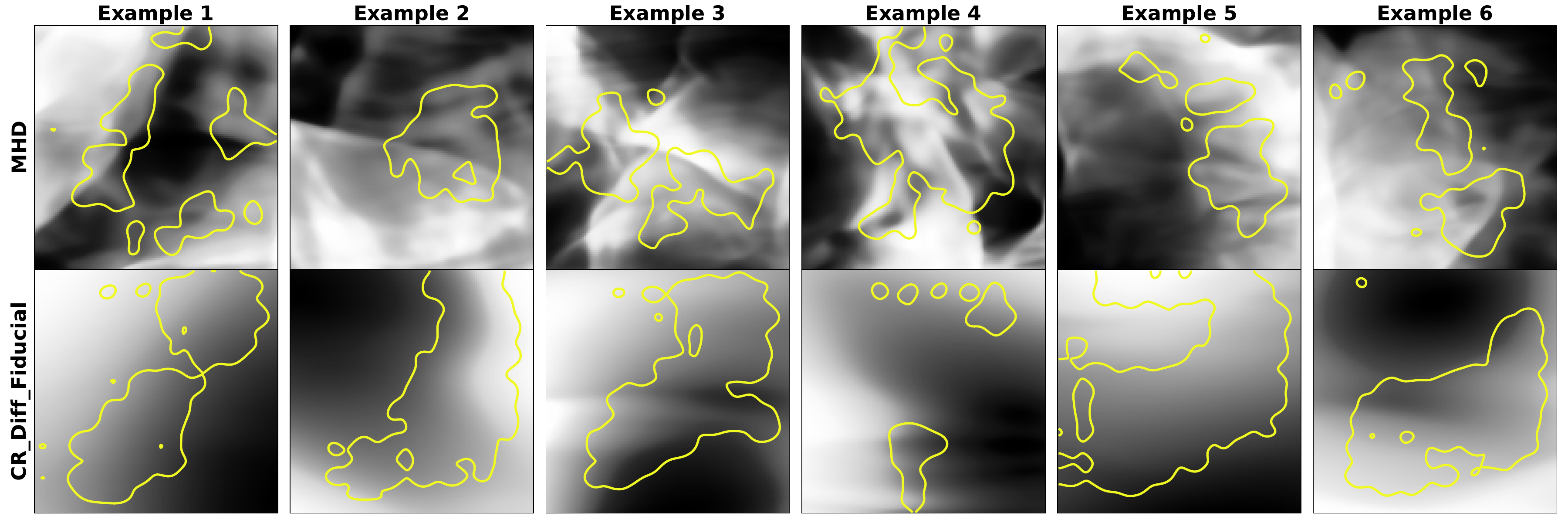}
\caption{Image examples and overlaid saliency contours for a neural network trained on just two classes: \texttt{MHD} (top row) and \texttt{CR{\_}Diff{\_}Fiducial} (bottom row). Saliency contours are informative here, showing that the \texttt{MHD} images are particularly distinguished by sharp edges and regions with lots of structure, while  the salient features of the \texttt{CR{\_}Diff{\_}Fiducial} images  are broad, smooth gray regions without much small-scale structure.} 
\label{fig:saliency_full_power_twoClasses}
\end{figure*}

We employ a combination of techniques to further interpret our results. First, we produce a set of saliency maps, which essentially show the regions of an image that produce large network activations (large gradients stored during backpropagation) leading to a final prediction. The first set of saliency maps, shown in Figure \ref{fig:saliency_colormesh}, is for the fiducial \texttt{Full Power} network trained on data from all 5 classes. Each column shows \emph{one} example image from each class in grayscale (the same example images in Figure \ref{fig:example_images}), and each row shows the activations from each class when presented with that image. These activations are overplotted with a white-to-red colormap. For improved readability compared to showing the pixel-by-pixel saliency, they are Gaussian filtered with kernel size = 16 and standard deviation $\sigma = 4$ (in units of number of pixels, meaning the kernel size and standard deviation are $L/4$ and $L/16$, respectively) using the \texttt{Scipy} \texttt{gaussian{\_}filter} routine. The density images themselves have not been Gaussian filtered. Note also that activations for each image are also normalized to the range [0,1]; this makes visualizing very small activations easier but misrepresents the relative magnitudes of activations for different classes.

While it's difficult to immediately see a trend, after looking at enough image sets, one can convince themselves that the most salient features of the \texttt{CR{\_}Diff{\_}Fiducial} images are the broad, diffuse gray regions most unique to that class. Instead, sharp transitions are apparent in each of the \texttt{MHD}, \texttt{CR{\_}Advect}, and \texttt{CR{\_}Diff100} images. Because the sharp features are not unique to any one class, they don't appear in our saliency maps. One might notice, however, that the \texttt{CR{\_}Diff{\_}Fiducial} saliency correlates well with gray regions.

It is hard to discern a trend that further distinguishes the \texttt{MHD}, \texttt{CR{\_}Advect}, and \texttt{CR{\_}Diff100}, and \texttt{CR{\_}withStreaming} classes despite applying saliency maps to a wide range of image sets, changing the color map, etc. One possible explanation is that the higher-level, distinguishing information is imprinted as a \emph{change in correlation over scales} rather than as a change to a local structure with well-defined boundaries, the former being very typical in physics and the latter being the typical use-case of saliency maps for e.g. object classification or detection \cite{gradcam, 2019arXiv190107683M}.

When we instead train a model with the same network architecture but only on the \texttt{MHD} and the \texttt{CR{\_}Diff{\_}Fiducial} images, network accuracy on the test set is $>99 \%$ for both classes (we omit a confusion matrix for brevity), showing the drastic changes imparted by diffusing CRs. Figure \ref{fig:saliency_full_power_twoClasses} shows 6 example images from each class and activations for the predicted class now shown as yellow contours of constant saliency. For example, images in the top row all belong to the \texttt{MHD} class, and all yellow contours in that row show the activations for the \texttt{MHD} prediction. Across all 12 images, it's quite obvious that what distinguishes the two classes is the sharpness of black and white transitions: smooth density transitions and regions without small-scale structure trigger \texttt{CR{\_}Diff{\_}Fiducial} predictions, while sharp features generally trigger the \texttt{MHD} predictions.

\begin{figure*}
\centering
\includegraphics[width=0.9\textwidth]{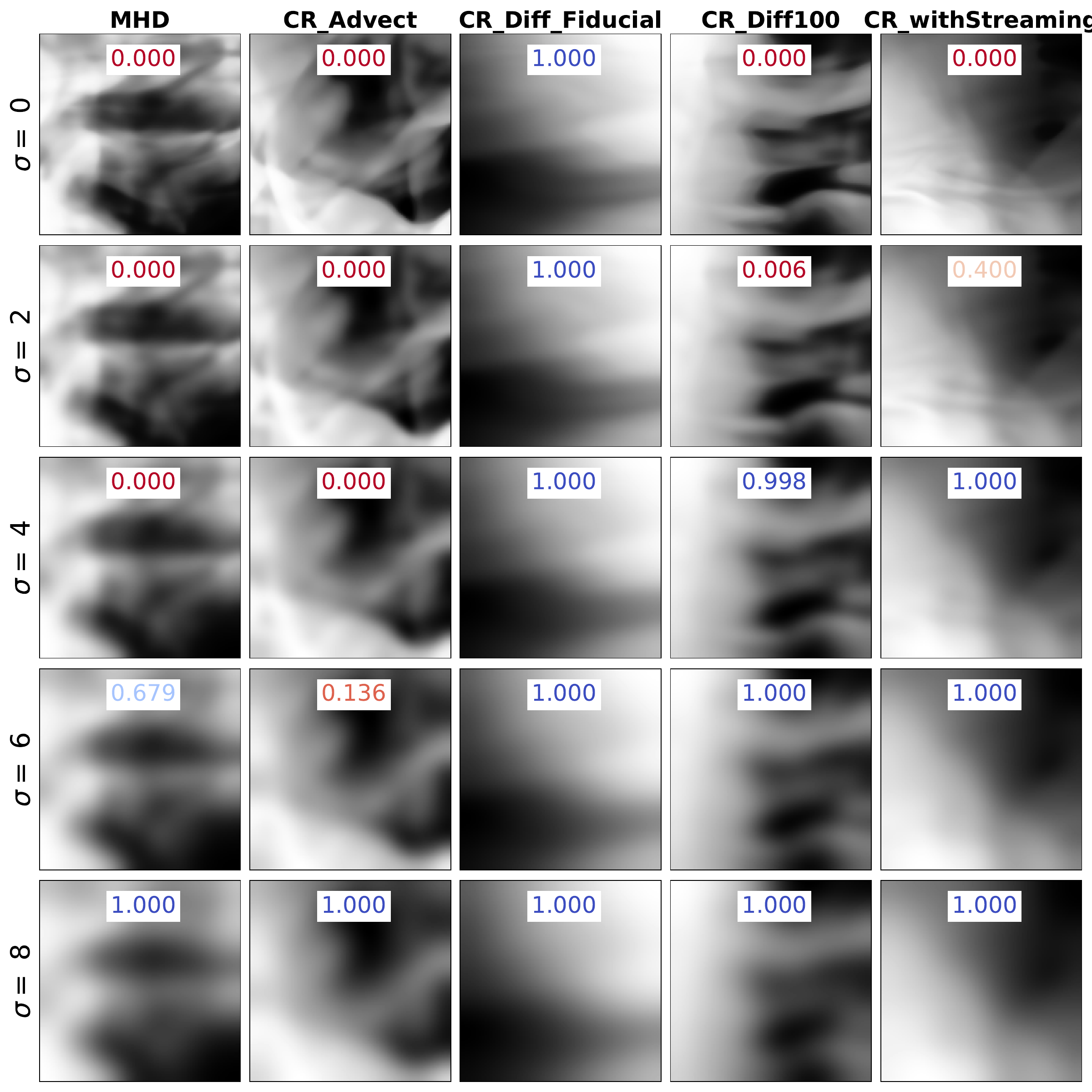}
\caption{Experiment whereby one image from each class (different columns) is Gaussian filtered to different degrees, parametrized by different standard deviations $\sigma$ of the Gaussian kernel. When run through the trained network, the output probability that the new image belongs to the \texttt{CR{\_}Diff{\_}Fiducial} class is denoted near the top of each image. Descending the rows (increasing $\sigma$), small-scale structure further disappears from each image, and for all classes, the images are eventually classified as \texttt{CR{\_}Diff{\_}Fiducial} with  $>99 \%$ confidence. } 
\label{fig:GaussianFiltering}
\end{figure*}

To help reveal the image features that led to predictions, we also see how the network handles a simple image manipulation: we take one test set image belonging to each class, and we Gaussian filter that image to varying extents by varying $\sigma$, the standard deviation of the Gaussian kernel. Figure \ref{fig:GaussianFiltering} shows the original images in the top row, followed by the more and more Gaussian filtered images going from top to bottom. On top of each image, we denote the network's probability that the manipulated image belongs to the \texttt{CR{\_}Diff{\_}Fiducial} class. As $\sigma$ increases, all probabilities converge to 1.0, showing that the \texttt{CR{\_}Diff{\_}Fiducial} images are, at least according to the network, the limit of significant ``filtering" due to CR-induced damping of small-scale features. 

\begin{figure}
\centering
\includegraphics[width=0.80\textwidth]{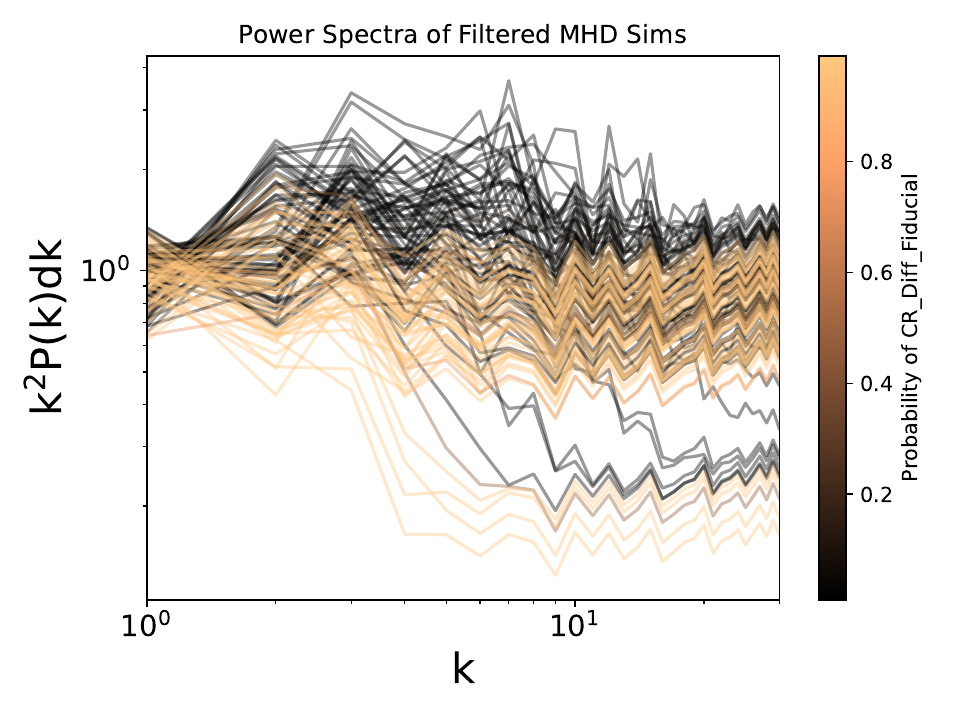}
\caption{Power spectra of a batch of \texttt{MHD} images Gaussian filtered with varying $\sigma$. Colors denote the probability that the network classified the image as \texttt{CR{\_}Diff{\_}Fiducial}. Note the delineating boundary around $P(k) \sim k^{-2}$; images with decreased power at small and intermediate wavenumbers $k < 10$ are confidently classified as \texttt{CR{\_}Diff{\_}Fiducial}, another sign that damping is a distinguishing feature of that class.} 
\label{fig:filtered_spectra}
\end{figure}

Figure \ref{fig:filtered_spectra} shows this in a different way. It shows the 1D power spectra (multiplied by $k^{2}$ to highlight differences) averaged for each class in a batch of 512 unfiltered test set images. Clearly, the \texttt{CR{\_}Diff{\_}Fiducial} class shows less power at intermediate scales than the \texttt{MHD} class, while the other CR classes have similar spectra. In particular, the \texttt{CR{\_}Diff100} and \texttt{CR{\_}withStreaming} classes have almost identical spectra, possibly leading to the confusion between those classes. The right panel shows the power spectra of \texttt{MHD} images filtered to varying extents. Unfiltered images typically have significant power at large and intermediate scales (small and intermediate wavenumbers, $k$) and are classified as \texttt{CR{\_}Diff{\_}Fiducial} with very low probability. Highly blurred images, however, lack as much power at $k < 10$ and are confidently classified as \texttt{CR{\_}Diff{\_}Fiducial}.

\subsection{Images with Flattened Spectra}
Motivated by \cite{Peek2019}, which showed that a trained CNN can distinguish sub-Alfv\'{e}nic vs super-Alfv\'{e}nic images even when density spectra are flattened (equaled), we flatten the power spectra of our images (the \texttt{Flattened Power} set) and train a separate classification network with identical architecture but different hyperparameters. This is especially interesting given what we have seen so far: that the network can learn the presence of (or lack of) spectral information, especially for the \texttt{CR{\_}Diff{\_}Fiducial} images, which show very little small-scale structure. However, as seen in Figure \ref{fig:original_spectra}, spectral differences between the other classes, namely the \texttt{MHD}, \texttt{CR{\_}Advect}, \texttt{CR{\_}Diff100}, and \texttt{CR{\_}withStreaming}, are quite small. This suggests the presence of other, non-spectral distinguishing features. By flattening the power spectra, we can probe this idea more explicitly and ask the network to find the salient \emph{phase} information in the images resulting from CR interactions with gas perturbations.

\begin{figure*}
\centering
\includegraphics[width=0.90\textwidth]{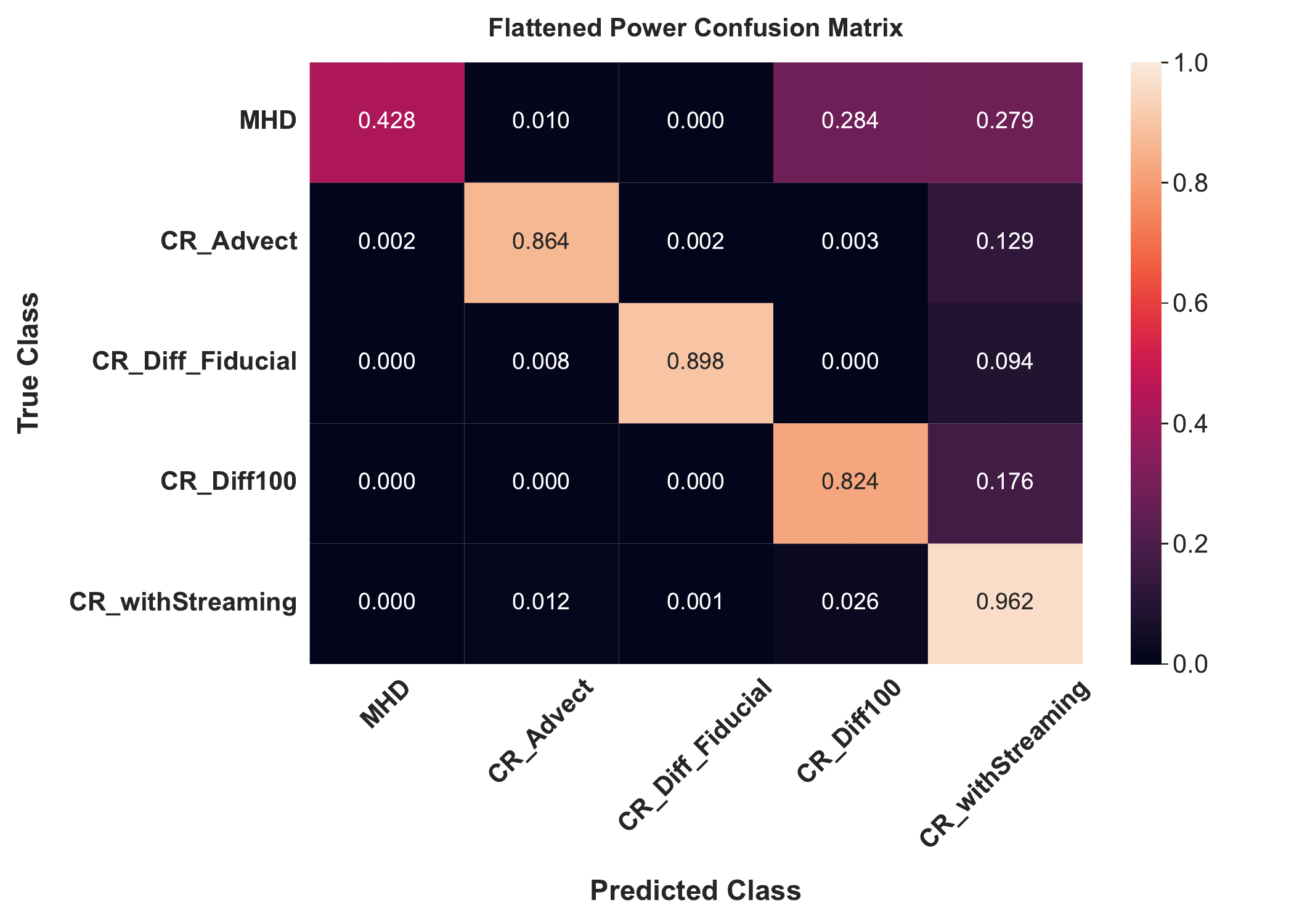}
\caption{Confusion matrix for the \texttt{Flattened Power} set of test images. Accuracies range from $\sim 85-98\%$; As all images have flattened power spectra now, clearly, CRs impart distinguishing \emph{phase} changes to the turbulent flow that the network appears to learn. The MHD case again is the least reliable, with a low recall of $42.8\%$ and frequent confusion with the \texttt{CR{\_}Diff100} and \texttt{CR{\_}withStreaming} classes. } 
\label{fig:confusion_kill_power}
\end{figure*}

\begin{figure*}
\centering
\includegraphics[width=0.8\textwidth]{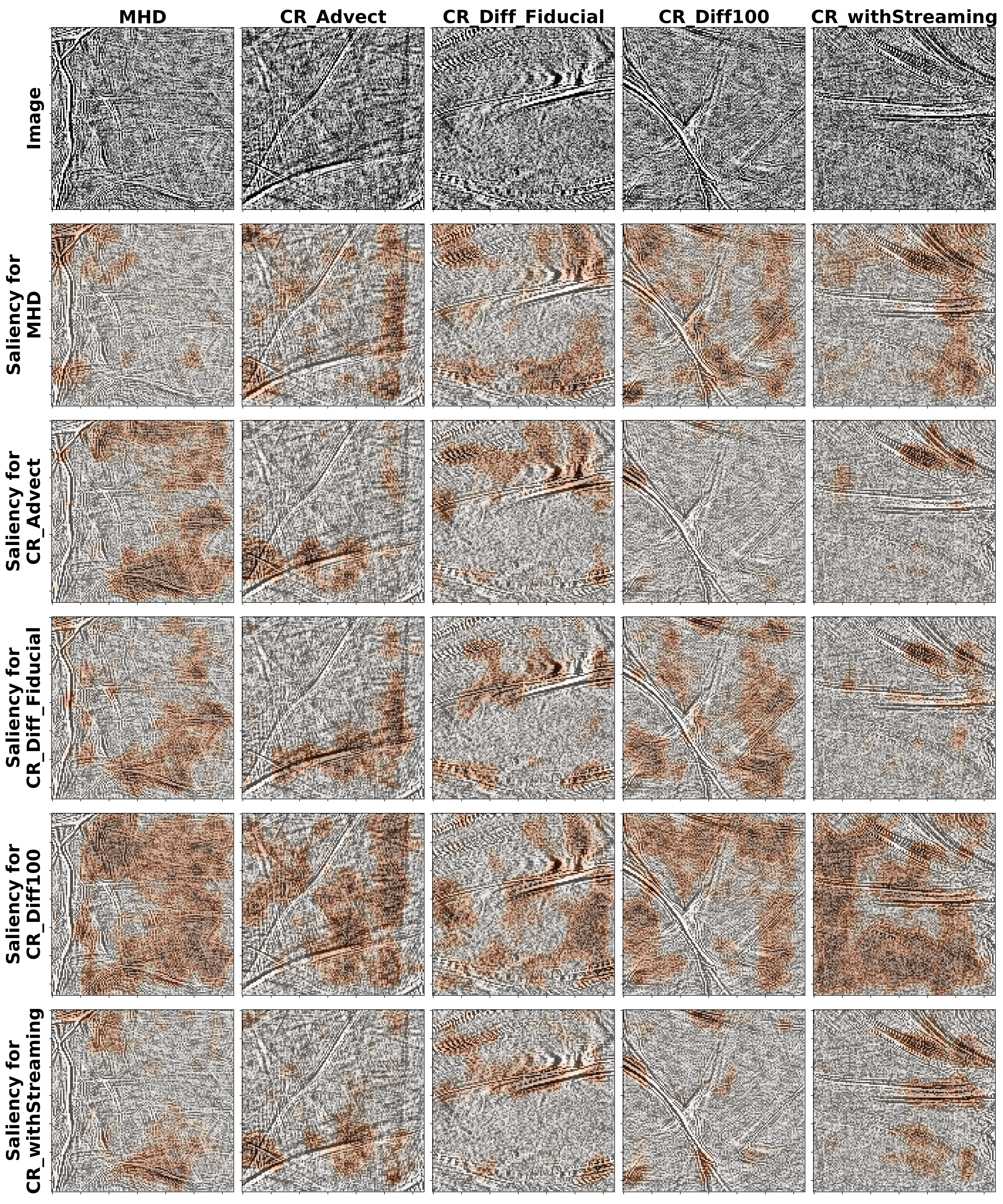}
\caption{Same as Figure \ref{fig:saliency_colormesh} but for example images from the \texttt{Flattened Power} test set. Many of the same regions are activated for each class; however, when looking at the activations for the true class, it's apparent in some cases that certain, isolated features drive the prediction. See e.g. \texttt{CR{\_}withStreaming}, in which case the true prediction map (``Saliency for \texttt{CR{\_}withStreaming}") shows a high activation in only one small area, whereas activations for other classes are more spread out.} 
\label{fig:saliency_kill_power}
\end{figure*}

The resulting confusion matrix for the test set is shown in Figure \ref{fig:confusion_kill_power}, and example \texttt{Flattened Power} images are shown in Figure \ref{fig:saliency_kill_power}. Class accuracies for the \texttt{Flattened Power} test set are comparable to those for the \texttt{Full Power} test set (see Table \ref{tab:metrics}), with \texttt{CR{\_}Diff{\_}Fiducial} again being the most accurately predicted class, although the recall has dropped from $100\%$ to $89.8\%$ in exchange for a higher precision of $99.7\%$. F1 scores amongst the CR classes are again quite high, reaching up to $94.5\%$ for \texttt{CR{\_}Diff{\_}Fiducial}. The most obvious change from the \texttt{Full Power} case is that the \texttt{MHD} and \texttt{CR\_withStreaming} F1 and accuracy scores dip significantly (by more than a few percent) when their spectra are flattened. Indeed the main differences between the confusion matrices (Figures \ref{fig:confusion_full_power} and \ref{fig:confusion_kill_power}) are the mistaken predictions of \texttt{CR\_withStreaming} when the true class is \texttt{MHD}. 

To reason why these summary statistics have changed, we must consider the physical, distinguishing characteristics that might exist even in the absence of spectral information. For instance, in sub-Alfv\'{e}nic turbulence, even with purely compressive forcing, spatial and spatial-temporal decompositions show that a significant portion of the energy lies in Alfv\'{e}n modes \cite{Makwana2020, Gan2022}, with solenoidal motions being generated by a combination of compressive motions and magnetic forces \cite{lim2020}. CRs, in a transport-dependent way, have been shown to affect the ratio of solenoidal energy $E_{\rm sol}$ to compressive energy $E_{\rm comp}$ and the scale-dependent mixture of these motions (see Section 4.5 of \cite{BustardOh2023_arxiv}). This in-turn influences the \emph{morphology} of density fluctuations: sharp, shock-like features indicate compressions and rarefactions, while ``swirls" indicate solenoidal motions.

Bustard and Oh 2023 measure $E_{\rm sol}/(E_{\rm comp} + E_{\rm sol}) \sim$ 0.42, 0.36, and 0.67 for the \texttt{MHD}, \texttt{CR{\_}Diff{\_}Fiducial}, and \texttt{CR\_withStreaming} classes, respectively. The increased fraction of solenoidal power in the \texttt{CR\_withStreaming} case should be well-imprinted in the \texttt{Flattened Power} image set, but the low precision of $58.7\%$ for \texttt{CR\_withStreaming} suggests otherwise. One possible reason is that the increase in solenoidal power is most acute at large scales $\sim L$ and almost negligible at smaller scales (see Figure 10 in \cite{BustardOh2023_arxiv}), meaning it might not be well-reflected in our images of size $L/2 \times L/2$. On the other hand, the \texttt{CR{\_}Diff{\_}Fiducial} class is well-separated from the others despite having an almost identical $E_{\rm sol}/(E_{\rm comp} + E_{\rm sol})$ to the \texttt{MHD} case. If the distinguishing image characteristics are due to solenoidal vs compressive motions, the network must be quite sensitive to them for some classes but not others. We caution, however, that even in the no-CR case, the mixture of modes in MHD turbulence is not well understood and is an active area of research. How CR transport further affects turbulent phase information and gas morphology is still a very open problem.

Alternatively, the confusion between \texttt{MHD} and \texttt{CR\_withStreaming} in the \texttt{Flattened Power} set but not the \texttt{Full Power} set may simply mean the differences are largely spectrum-related rather than phase-related. As evidenced by Figure \ref{fig:filtered_spectra}, there \emph{are} some small differences between the image spectra that may have been critical in the \texttt{Full Power} case but have now been thrown out in our \texttt{Flattened Power} study.

To try to interpret our network, we again employ saliency maps on a set of example images in Figure \ref{fig:saliency_kill_power}. By eye, the \texttt{MHD} and \texttt{CR{\_}Advect} image sets show long narrow features indicative of strong compression fronts, while the \texttt{CR{\_}Diff{\_}Fiducial} image lacks this sharp structure. However, these differences are not necessarily activated by our saliency experiment, as was the case for our \texttt{Full Power} data set. The \texttt{CR{\_}withStreaming} activations, however, consistently line up with strong line features across several example images from different classes. That these are seemingly such strong indicators of streaming CR transport is encouraging in our quest to determine the true CR transport mode in different astrophysical environments; however, these features also bear strong resemblance to the compression fronts in the \texttt{MHD} images, and significant confusion between those classes may be related to these similar features.

We experimented with a few other interpretability tools, most notably occlusion experiments where one asks the network to predict an image with obscured regions (ideally regions where salient features are present), but no experiments to-date have gleamed much new information. For brevity, we stop our exploratory analysis here, but follow-up studies can probe the origin of distinguishing, non-spectral features, which are possibly related to the CR transport-dependent mixture of solenoidal and compressive motions found in Bustard and Oh 2023. Related work using wavelet scattering transforms, which construct similar representations as CNNs (e.g. \cite{bruna2013invariant,Cheng+2023,Velicheti+2023}), has demonstrated the importance of encoding phase information and scale separation. Additional image manipulations and explicitly adding image spectra as an extra input to the \texttt{Flattened Power} model would be useful next steps to reveal and isolate additional distinguishing features and gain deeper insights on CR-induced differences.

\section{Limitations and Future Work}
\subsection{Simulation Limitations}
\label{sec:sim_limitations}

While our results are promising, this method is not without its limitations. The biggest issue we hope to tackle in future work is domain adaptation: can this network, trained on simulation data, generate accurate predictions when deployed on real observations? For this supervised learning algorithm to back out CR transport from observations, we require the ``ground truth" given to us by simulations, but the simulations have a number of restrictions that could limit our machine learning model's ability to generalize to observations: 
\begin{enumerate}
    \item {\bf Physics choices:} For instance, these simulations use an isothermal equation of state instead of an adiabatic equation of state with realistic radiative cooling, conduction, self-gravity, etc. Additionally, ensuring the correct ionization state of the gas is crucial to making a robust connection to observations, but this is very environment-specific and dependent on the local distribution of nearby massive stars, etc.
    \item {\bf Simulation parameter coverage:} For instance, the results shown here are trained on a simulation suite with a number of parameters fixed -- the stirring rate, the initial plasma beta, etc. One could increase the span of the training data across different parameters, but the data volume would quickly become very large. 
    \item {\bf Simulation convergence:} Bustard and Oh 2023 conducted a limited convergence study, showing that the main transport-dependent trends (CR-induced damping of small-scale fluctuations, etc.) are robust to changes in resolution; however, increasing resolution will always lead to more small-scale structure because the turbulent inertial range, which artificially dissipates in simulations on length scales of $\approx 30$ cell widths \cite{Federrath2010}, extends to smaller scales. Higher resolution simulations, being more computationally intensive, are not possible at this point, but in the future, one might create a more robust network by training it on simulations with varying resolution.  

\end{enumerate}

One could also map simulations to observations more closely by folding in additional telescope effects during preprocessing, but this depends on the case-by-case telescope instrumentation and is beyond the scope of this paper.

\subsection{Sensitivity to Projection Depth} 
\label{sec:appendix}

Alleviating the issues above will require additional simulations. With our existing simulation suite, though, we can quickly explore another limitation: our use of single-cell-thick slice plots as training data, rather than projections that integrate further along our line-of-sight. To test this, we create two additional image sets by averaging over $d$ cells perpendicular to the image plane, rather than creating images from single-cell slices. The resulting test sets have roughly $3,800$ and $760$ images per class for $d = 8$ and $32$, respectively. 

\begin{figure*}
\centering
\includegraphics[width=0.49\textwidth]{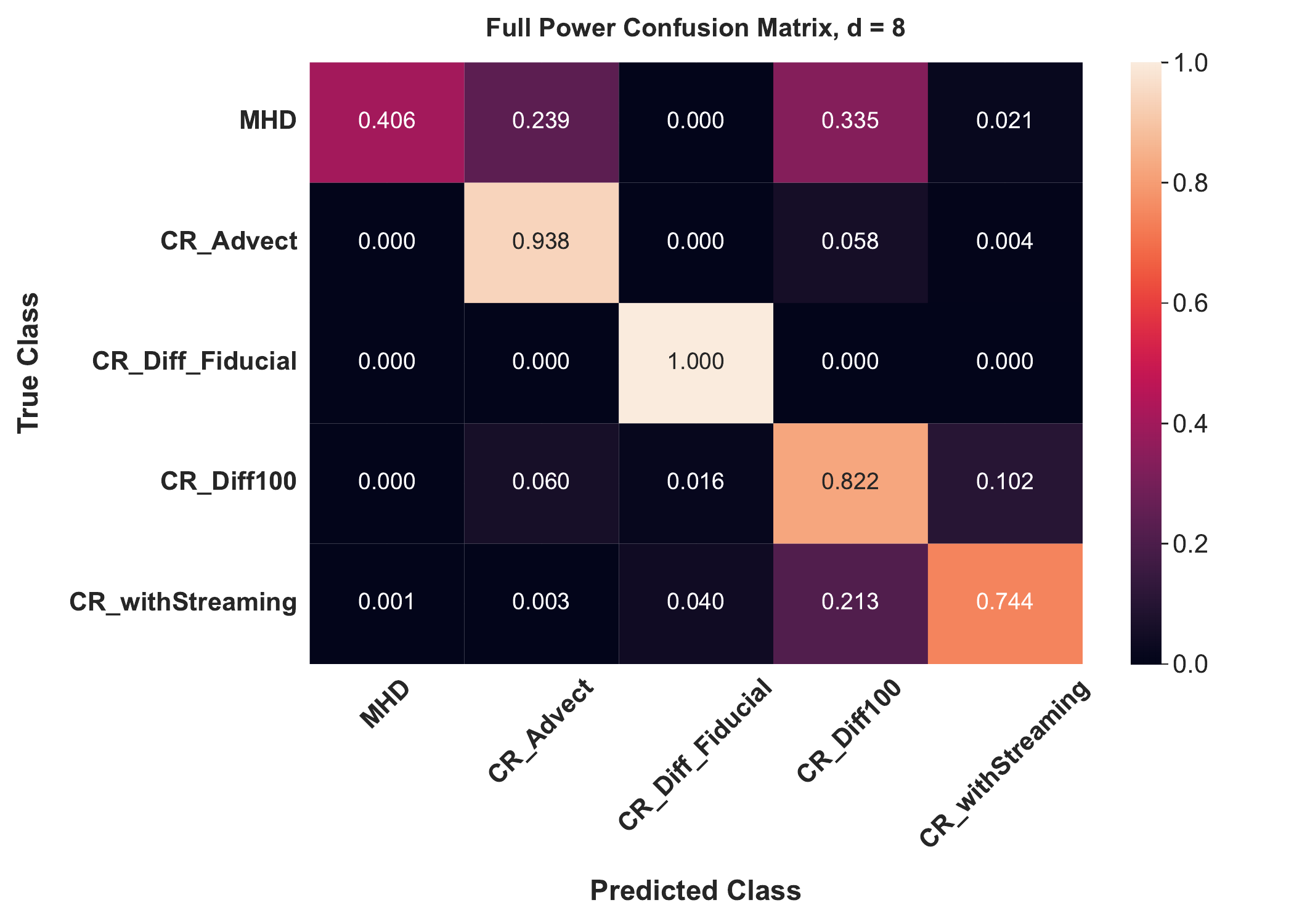}
\includegraphics[width=0.49\textwidth]{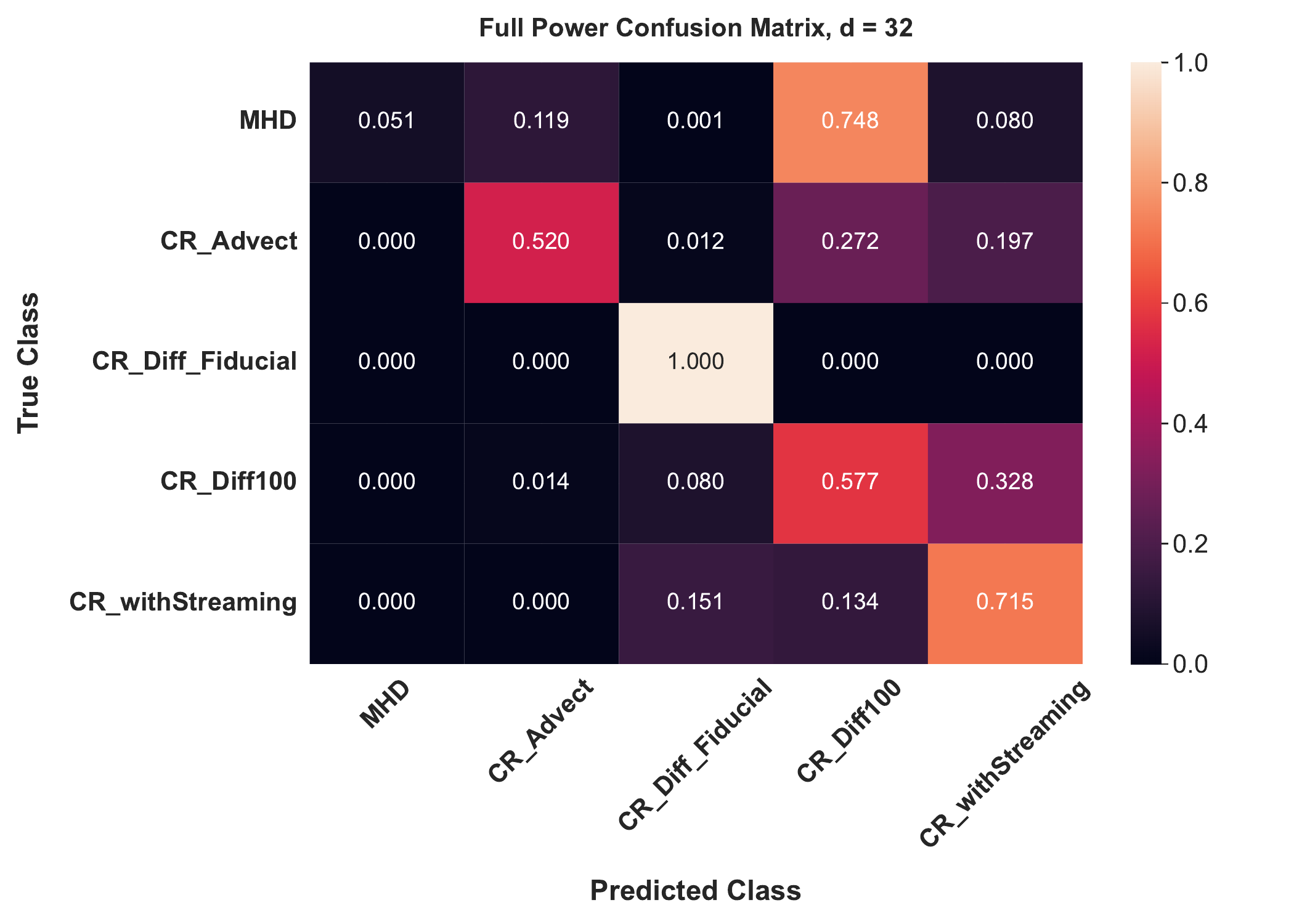}
\caption{Confusion matrix for our pre-trained \texttt{Full Power} model applied to images created by averaging over $d = 8$ cells (left) and $d = 32$ cells (right) along the $\hat{x}$-axis. Note how accuracies decrease significantly, particularly for classes that originally were characterized by sharp density gradients and small-scale structure; these salient features are smoothed out when averaging over more and more layers, leading to numerous misclassifications.} 
\label{fig:proj_depth8}
\end{figure*}

Figure \ref{fig:proj_depth8} shows confusion matrices when our model (pre-trained on slices) is asked to predict images with $d = 8$ and $32$. In both cases, the model can very accurately classify the \texttt{CR{\_}Diff{\_}Fiducial} images, but as the depth increases, overall accuracy decreases significantly. This is most true for the \texttt{MHD} images, which show the crux of the issue: averaging / projecting over multiple layers smooths out the small-scale structure that distinguishes the other classes, particularly the \texttt{MHD} class, from the others.

If there is a significant gap between the training projection depth and the test projection depth, then accuracy can degrade significantly. This is not entirely unexpected, since this ``model misspecification'' or ``domain shift,'' i.e. applying a model trained on one dataset to another, is an active and unsolved area of research (although there have been recent promising results in domain adaptation for astronomical machine learning; e.g. see \cite{ciprijanovic_deepastrouda}).

It is possible to increase our accuracy by training a model specifically on images with the same projection depth as in the validation set, thereby bypassing the model misspecification problem. However, this is unlikely to be helpful in real observations of turbulent gas clouds, which are subject to uncertain distance measurements \cite{Green+2019}. Moreover, we do not have strong constraints on how far we can see into the gaseous structure (i.e. the optical depth), which means that we cannot estimate \textit{a priori} the number of turbulent eddies that we would expect to see in the line-of-sight direction and, therefore, cannot determine an appropriate simulation projection depth $d$ to compare to.

Overall, given the critical differences between current simulations and real observations, challenges associated with observational uncertainties, and the problem of domain shift in supervised machine learning, we caution against directly applying our method to real data. However, our publicly available code can serve as a useful framework for data preprocessing and model training, while our publicly available data volumes, which are scale-free and can therefore be scaled up or down to different astrophysical regimes depending on the turbulent driving scale $L_{0}$, can serve as a useful comparison to more detailed simulations in the future.

\section{Conclusions}
\label{sec:conclusions}

Deducing CR transport physics from observations is a massive undertaking, involving phenomenological models fit to direct and indirect CR observables \cite{Hanasz2021}, galaxy and zoom-in cosmological simulations compared to radio synchrotron and gamma-ray emission \cite{Chan2019, hopkins2022-problems}, and focused probes of CR penetration in cold molecular clouds \cite{Everett2011, Morlino2015, Dogiel2018, Bustard2021} and CR transport along radio-emitting filaments \cite{Thomas2020RadioHarps}. All of these research avenues rely on multi-wavelength data, e.g. high-energy emission from radio emitting CR electrons or gamma-ray emission arising from hadronic interactions of thermal gas with CR protons. An alternative, and to our knowledge unexplored, avenue is to harness recent advances in deep learning and a growing amount of simulation data to train a network to recognize CR transport physics from solely density images. 

In this paper, we trained and fine-tuned multi-layer convolutional neural networks (CNNs) on a suite of turbulent box simulations \cite{BustardOh2023_arxiv} with varying CR transport prescriptions from pure CR advection to CR diffusion to CR streaming. We also use interpretability tools like saliency maps and image manipulation to interpret these results and to help build physical intuition for CR impacts on turbulence.

The main findings of this work are: 
\begin{itemize}
    \item Our trained CNN can classify images originating from simulations with 5 different CR transport prescriptions with high class accuracies ranging from $92.0\%$ to $99.2\%$ and F1 scores ranging from $80.8\%$ to $98.0\%$. The average recall is brought down most significantly by the MHD-only (no CR) simulations, whose resulting density images closely resemble those with either very little or very fast CR transport. 
    \item Images derived from simulations with intermediate diffusivity, i.e. \texttt{CR{\_}Diff{\_}Fiducial}, are most accurately classified ($99.2 \%$). Saliency maps (Figures \ref{fig:saliency_colormesh} and \ref{fig:saliency_full_power_twoClasses}) identify smooth, rather than sharp, density contrasts as distinguishing features of these images, owing to strong CR-induced damping of small-scale turbulent fluctuations \cite{BustardOh2023_arxiv} that effectively Gaussian filters or ``blurs" the image (Figure \ref{fig:GaussianFiltering}). 
    \item Streaming and diffusion lead to distinctly different images and are only rarely confused by our trained network (Figure \ref{fig:confusion_full_power}); however, there is some confusion between streaming transport and fast diffusive transport. 
    \item Images with flattened power spectra are also classified with high accuracies (85.7-97.9\%), suggesting that CRs change both spectral \emph{and} phase information, as would be the case if CRs affect the balance between compressive and solenoidal motions as found in \cite{BustardOh2023_arxiv}. In particular, saliency maps (Figure \ref{fig:saliency_kill_power}) reveal that the network consistently associates streaming transport with strong lines in spectrally flattened gas density images.
\end{itemize}

\ack
The authors gratefully acknowledge Peng Oh, Josh Peek, and Blakesley Burkhart for stimulating discussions, as well as the organizers and participants of the KITP program ``Building a Physical Understanding of Galaxy Evolution with Data-driven Astronomy" where this work originated. This research was supported in part by the NSF PHY-2309135 grant to the KITP. CB was supported by the National Science Foundation under Grant No. NSF PHY-1748958 and by the Gordon and Betty Moore Foundation through Grant No. GBMF7392. Turbulence simulations were performed on the Stampede2 supercomputer under allocation TG-PHY210004 provided by the Extreme Science and Engineering Discovery Environment (XSEDE), which is supported by National Science Foundation grant number ACI-1548562 \cite{xsede}.

\section*{References}
\bibliography{bibliography}

\end{document}